# So, I Can Feel Normal: Participatory Design for Accessible Social Media Sites for Individuals with Traumatic Brain Injury

Hajin Lim*

Department of Computer Sciences, University of Wisconsin-Madison, United States, hajin@snu.ac.kr

Lisa Kakonge

Rehabilitation Science, McMaster University, Canada, kakongel@mcmaster.ca

Yaxin Hu

Department of Computer Sciences, University of Wisconsin-Madison, United States, yaxin.hu@wisc.edu

Lyn S. Turkstra

Rehabilitation Science, McMaster University, Canada, turkstrl@mcmaster.ca

Melissa C. Duff

Department of Hearing and Speech Sciences, Vanderbilt University Medical Center, United States, melissa.c.duff@vanderbilt.edu

Catalina L. Toma

Department of Communication Arts, University of Wisconsin-Madison, United States, ctoma@wisc.edu

Bilge Mutlu

Department of Computer Sciences, University of Wisconsin-Madison, United States, bilge@cs.wisc.edu

Traumatic brain injury (TBI) can result in chronic sensorimotor, cognitive, psychosocial, and communication challenges that can limit social participation. Social media can be a useful outlet for social participation for individuals with TBI, but there are barriers to access. While research has drawn attention to the nature of access barriers, few studies have investigated technological solutions to address these barriers, particularly considering the perspectives of individuals with TBI. To address this gap in knowledge, we used a participatory approach to engage 10 adults with TBI in conceptualizing tools to address their challenges accessing Facebook. Participants described multifaceted challenges in using social media, including interface overload, social comparisons, and anxiety over self-presentation and communication after injury. They discussed their needs and preferences and generated ideas for design solutions. Our work contributes to designing assistive and accessibility technology to facilitate an equal access to the benefits of social media for individuals with TBI.

---

* Current Affiliation: Department of Communication, Seoul National University, South Korea

# 1 INTRODUCTION

Traumatic brain injury (TBI) is brain damage caused by external force that results in altered brain function [15,27]. The incidence of TBI in adults is estimated at nearly 1.5 million per year in the U.S alone, with 80,000 to 90,000 people experiencing long-term, chronic disability after their injuries [27]. Effects of TBI can include chronic physical, sensory, cognitive, and communication impairments [15]. Challenges in social communication are particularly common, including difficulty initiating social interactions and conversing with others [61,67]. These challenges contribute to the loss of significant relationships and friendships that is the hallmark of moderate-severe TBI and is often accompanied by feelings of isolation and depression [36,42,97]. Changes in maintaining and building social networks [97] and limited opportunities for socialization [29] may negatively affect not only face-to-face interactions of individuals with TBI, but also social media interactions [18,54].

Research on uninjured populations has identified many social benefits stemming from social media use. In general, communicating with "friends" or followers on social media has been associated with increased social capital [41,108] and civic engagement [116], reduced loneliness [35], and increased enactment of relationship maintenance behaviors [41]. Additionally, broadcasting information about oneself on social media can produce self-affirmation [98] and boost self-esteem [46], as well as increase positive affect among users [29]. These functions and benefits of social media for people with TBI are well documented [8,18,22,72] as well For example, people with TBI report using social media to connect with others who share similar experiences [18], maintain relationships with close friends and family , and make new friends [72], gather health information [18,72], and advocate for TBI-related causes [19]. Successful social media use for these varied functions may help reduce feelings of isolation and improve overall well-being [3]. For example, one study of adults with TBI found that social media use was related to higher levels of self-reported life satisfaction [42].

While social media can offer opportunities for social participation and psychological well-being, social media use also comes with general drawbacks, such as excessive and compulsive use [57], negative feelings and social comparisons [96,104,111], or fatigue [16]. Social media use, and computer mediated communication more broadly, can be particularly challenging for individuals with TBI, due to their sensory and cognitive impairments [45]. In particular, the sheer volume of content, complexity of the interface, and information overload and the necessity to multitask [45] may contribute to exacerbated cognitive fatigue [22,92]. Also, because of their social communication challenges, people with TBI also might be at risk for Internet scams, cyberbullying, misinformation, and privacy violations [18]. These TBI-related challenges may explain, in part, reports of reduced social media use after injury [8,72] and may prevent individuals with TBI from deriving the full benefits of social media participation. Considering the general drawbacks of using social media can be more pronounced with TBI-related challenges, it is important to ensure that individuals with TBI have the same access to the benefits of social media as those without TBI, while minimizing the disadvantages and challenges associated with TBI.

However, there is still a lack of consideration for supporting the social and communication needs that individuals with TBI may have using social media. Until now, research has focused on identifying challenges that individuals with TBI face using social media rather than identifying solutions for their challenges. Further, social media sites have primarily offered accessibility options to support people with auditory and visual limitations, such as alt-text, video captions and color contrast adjustments [45,110]. By contrast, there has been relatively little attention to improving social media accessibility for individuals with cognitive disabilities, including those with TBI [14,23,72].

The purpose of this study was to learn how adults with TBI used social media and elicit their ideas and feedback about technological tools that would help them use social media as wished. We conducted remote participatory design (PD) sessions with 10 individuals in the chronic stage after TBI. We first conducted the exploration activity that involved interviews and think-aloud sessions to better understand participants' challenges with social media sites, focusing primarily on their lived experiences with Facebook, one of the social media sites most commonly used by individuals with TBI [72]. Next, we conducted ideation and sketch activities in which participants engaged in ideating solutions to address the challenges they identified. Through PD sessions, we identify four main themes of TBI that include: "interface, feature, and information overload", "Facebook as a source of negativity and social comparisons", "self-presentation concerns around TBI", and "reduced perceived confidence in communication." We share the solution ideas that participants came up with to address these challenges.

The primary contributions of our work include: (1) expanding the research body on challenges to social media usage after TBI and (2) presenting solutions envisioned by individuals with TBI, shedding light on the future design of social media sites and assistive and accessibility tools for individuals with TBI. Not only do these solutions provide insight into individuals with TBIs' visions for themselves, but also have implications for the improvements to accessibility for broader populations that face similar challenges. Additionally, the considerations and procedures we designed and implemented into our remote PD method can be applied to future studies that actively engage individuals with TBI in their PD process.

## 2 RELATED WORK

We focus our related work on three major categories: (1) understanding general TBI symptoms and their impact on social media uses, (2) examining assistive and accessibility technology for individuals with TBI, and (3) discussing the value and consideration for running PD sessions with people with cognitive challenges including those with TBI.

### 2.1 Living with TBI

TBI can result in chronic deficits in cognitive and communication functions that are critical for accessing social media. These include impairments in learning and recall of previously learned information; selecting, and focusing attention; social cognition, and executive functions [9,25,68]. All of these cognitive functions are critical for navigating social media. In particular, social cognition includes the ability to "read" others' emotions and thoughts [25], including social cues in emojis and written text. Executive functions play a crucial role in task-oriented behaviors, including organizing, planning, prioritizing, and monitoring tasks [94], and navigating and using social media sites [100]. Cognitive impairments in turn contribute to deficits in aspects of interpersonal communication such as difficulty understanding information, problems finding words or being unable to understand what others are saying, inability to express thoughts and feelings clearly, and, poor organization of thoughts, ideas, and words [25,61]. Effects of TBI on social communication also may include problems initiating social interactions or conversing with others [61] and using context-appropriate language [25]. Further, impairments in basic sensory processes [2], including blurred vision and sensitivity to light [60] also can result in discomfort when using digital screens [62].

TBI can also have profound effects on mental health, from depression and anxiety [73,85] to post-traumatic stress disorder (PTSD) [103], which further presents challenges to social participation. Anxiety and depression can have a reciprocal relationship with decreased opportunities for friendships and social support or establishing new social contacts [73], so that the person becomes more socially withdrawn over time. Changes in personal abilities, decreased social contact, and reduced vocational options can lead to an altered sense of identity [113]. These challenges can translate into changes in online communication following TBI [42], ultimately limiting the ability of individuals with TBI to capitalize on the opportunities for connection offered by social media. Despite these TBI-related challenges, previous research suggests that individuals with TBI maintain social media accounts to the same extent as healthy comparisons [72,100] and are highly interested in using social media for social connection [20,22]. The literature also highlights the importance of using social media as a tool for social participation [19,23,24,86], which supports the goal of increasing social media access for individuals with TBI. [72]

## 2.2 Assistive Technology and Accessibility Options for individuals with TBI

Research in Human-computer Interaction (HCI) and rehabilitation has focused on providing training tools to people with TBI to enhance their cognitive, sensorimotor, and related living skills. One category of assistive tools provides cognitive support for the memory and attention deficits that tend to follow TBI, which include in-vehicle assistive tools that promote safe driving [80], mobile applications that support meal preparation [84], and tools that provide support in carrying out multi-step tasks [56]. These assistive tools were designed to help individuals with TBI plan and carry out multi-step tasks by prompting them to take the necessary actions. Another category of assistive tools has focused on supporting motor recovery through motion games. For instance, Cheng and colleagues [28] proposed game design patterns for integrating motion-based games into TBI therapy to enhance patients' motor abilities, including balance and weight shifting. Many of these existing tools, however, tended to only involve individuals with TBI in evaluation of developed prototypes by researchers (e.g., [56,80,84]). As such, it is relatively rare to involve individuals with TBI in the early ideation stage in the user-centered design cycle, which may limit the possibilities for identifying and addressing a variety of challenges and needs that they may face in their everyday lives [44,50].

A gap also exists in the accessibility options for supporting the social and communication needs of individuals with TBI [19,53]. For example, although social media sites have been striving to include more accessibility features in their platforms, those have mostly been geared towards people with hearing and vision impairments [45,110]. In contrast, there has been relatively little attention paid to improving accessibility for individuals with cognitive disabilities, including those with TBI [14,72]. While Web Content Accessibility Guidelines (WCAG) do include the requirements for cognitive accessibility [106] and some research offers general guidelines on interface design for individuals with TBI [30,77], many social media platforms have not yet fully integrated these principles and do not offer accessibility options tailored to the wide array of challenges faced by individuals with TBI. The lack of assistive technology and accessibility options for social media platforms may negatively impact their ability to capitalize on the opportunities for connection offered by these media.

## 2.3 Participatory Design and Individuals with Cognitive Impairment

Participatory design (PD) is a research methodology that involves participants as an essential part of the design process and bridges the tacit knowledge of participants with lived experience with theoretical knowledge of researchers [95]. PD views users as experts who are knowledgeable about their experiences and the context in which they live [89]. The goal of PD is to fully engage people in finding solutions to personally relevant problems. A large body of literature adopted PD approach to articulate their understandings and opinions about possible solutions to problems facing communities or groups (e.g., [75,88,89]).

However, research and design challenges can arise when using PD with people with cognitive or sensory impairments [44,50]. PD techniques often require higher-order cognitive skills, such as creative thinking, problem solving, and decision making [44], which could be challenging for individuals with TBI-related cognitive impairments [52]. The nature and severity of participants' individual cognitive impairments also may necessitate modifications to the PD procedure. Participants who live with cognitive or sensory impairments, for example, might need more time to consider their options or respond to a design challenge before they can act on it, which may require adaptations in how the researchers approach a task [71]. Accordingly, several researchers have proposed guidelines to make PD processes more accessible for people with cognitive impairments to accommodate their cognitive, communication, and environmental needs. To provide cognitive and communication support to participants with cognitive impairments, Mohatt and colleagues [71] emphasized the importance of providing instructions in a clear manner; repeating information as necessary and requested; using multiple modalities; and documenting the discussion in real time to facilitate participant recall. They also note that having necessary materials available to review before the study session help them "get in the right mindset" (p. 4). Furthermore, they discuss the importance of environmental considerations that include adequate space for PD sessions, ambulation, and natural lighting. Along with these considerations, as PD has been conducted in remote settings with the COVID-19 pandemic, the importance of technical support and rapport building processes among remote participants and researchers have also been emphasized [31,63]. In this study, we iteratively designed our PD sessions for participants with TBI based on these considerations, as described next in the Methods section.

## 3 METHODS

We conducted remote PD sessions to better understand the challenges that individuals with TBI face using social media (particularly Facebook) and to learn their design solutions for assistive and accessibility tools that can better support their social media use.

We do recognize that the vital components of PD include face-to-face group interactions, trust building, and collaborative group thinking [93]. Therefore, following the conventional PD process, we initially considered running the sessions in person, grouping two or three participants together. However, difficulties in recruiting local participants during the COVID pandemic lockdown made this consideration almost impossible. After we decided to run the sessions virtually, we revised the original study protocol to accommodate the remote PD setting on the basis of best practices for remote PD sessions [31,63] (see **Table 2**). Through multiple pilots and review sessions over the revised protocol, we came to conclude that one-on-one PD sessions would be more appropriate for our study population. Previous research suggested that individuals with TBI often exhibit impaired self-regulation of behavior [53], which can contribute to shifts in group dynamics in PD activities [26] (e.g., one individual dominates group discussion; or, conversely, demonstrates a lack of initiative to contribute, both of which are common behaviors among adults with TBI). Therefore, we saw that a one-on-one setting would be more ideal for providing personalized assistance and accommodation to participants while addressing their possible cognitive and communication challenges. As we turned to one-on-one PD sessions, we further envisioned the ways to build trust and rapport between individual participants and the facilitator as compensation for the group process. In addition, while running PD sessions with our participants, we adjusted our protocol iteratively as we encountered unexpected issues. We describe the finalized protocol in detail in the following procedure sections.

### 3.1 Research Team and Stance

Our study team was composed of HCI researchers interested in developing interactive technologies to facilitate inclusive online communication, and researchers with extensive clinical experience providing rehabilitation services to individuals with TBI. Though we have different backgrounds, we share a common belief that individuals with TBI were entitled to and could also have an equal access to the benefit of social media. Also, as a group, we strongly believed that individuals with TBI could actively collaborate with researchers to create knowledge about social media experiences and ideas for technical aids that could help their own online social participation. It was also advantageous to us to have different expertise in each field in order to compensate for each other's knowledge gaps in our respective fields of study. Researchers from the HCI field had knowledge of the PD method itself and translating outcomes from PD sessions into design implications but had little experience interacting with and accommodating individuals with TBI appropriately. Thus, our researchers from the rehabilitation field ran the PD sessions using their expert knowledge and experience in brain injury rehabilitation to build rapport and trust with individuals with TBI.

While we believe that this combination of expertise provides a solid foundation for developing supports to make social media accessible to individuals with TBI, we acknowledge that our shared characteristics (e.g., non-TBI individuals, educated, high technology literate, have had access to new technology) may prevent us from truly understanding and capturing their unique experiences.

### 3.2 Participants

For recruitment of participants, we distributed a recruitment poster to TBI rehabilitation service providers, posted it on a Facebook-group page of TBI service providers, and Reddit (r/TBI) and contacted one of the brain injury patient registries in the U.S. Individuals who expressed interest were invited to a pre-screening phone call to assess their eligibility for participation. Participants were required to have a diagnosis of TBI, be age 18 years or older, have a Facebook account, and be their own legal guardian for the purpose of consenting to research participation.

A total of ten adults with TBI (5 males, 5 females) participated in this study. Median participant age was 42 (range = 31-60 years). Despite efforts to recruit a racially diverse sample, all participants were white. The most common cause of TBI was a motor vehicle accident, followed by falls and sports injuries. At the time of PD sessions, a median of 7.5 years had elapsed since participants' first TBI (range = 3-42 years). Participants reported a variety of self-reported impairments related to their TBI, as noted in **Table 1**.

Table 1: Participant demographic and background information

| P | Age | Sex | Years post TBI | Sensory/physical challenges (e.g., vision, fatigue) | Cognitive challenges (e.g., memory, attention) | Psychosocial challenges (e.g., anxiety, mood swing, depression) | Language challenges (e.g., speech comprehension and production) |
|---|-----|-----|----------------|-----------------------------------------------------|------------------------------------------------|-----------------------------------------------------------------|-------------------------------------------------------------------|
| 1 | 37 | M | 7 | ✓ | ✓ | ✓ | ✓ |
| 2 | 50 | F | 42 | ✓ | ✓ | ✓ | ✓ |
| 3 | 60 | F | 5 | ✓ | ✓ | ✓ | |
| 4 | 32 | M | 6 | | ✓ | ✓ | |
| 5 | 32 | M | 3 | ✓ | ✓ | ✓ | ✓ |
| 6 | 37 | F | 17 | ✓ | ✓ | ✓ | ✓ |
| 7 | 53 | M | 18 | | ✓ | ✓ | |
| 8 | 32 | M | 10 | | ✓ | ✓ | |
| 9 | 31 | F | 5 | ✓ | ✓ | ✓ | ✓ |
| 10 | 56 | F | 8 | ✓ | ✓ | ✓ | ✓ |

### 3.3 Procedures

Before the scheduled session, we provided participants with a detailed overview of the PD activities and session reminders through phone calls, texts, and emails. We also inquired about participants' specific needs, preferences, and technological skills so that we could prepare adequate accommodations and supports ahead of time for each participant.

PD sessions were conducted individually with the researchers remotely over Zoom and were divided into three central activities: 1) exploration, 2) ideation, and 3) sketching, which all together took about 2 hours. Throughout the session and for each activity, we provided diverse strategies and accommodations that included communication, cognition, technology and input support outlined by literature [31,63,71], as presented in **Table 2**. All procedures were approved by ethics institutional review boards from the two universities where this study was hosted.

Table 2: Types of Accommodations provided in the PD Session

| PD Stage | Types of Accommodations (communication[1]; cognition[2]; technology[3]; input[4]) |
|---|---|
| **Pre-PD** | - (In pre-screening interview) Provided detailed introduction about activities and procedures in the session and introduced kinds of accommodations participants may want for virtual PD session[2]<br>- (Before scheduled PD session) Inquired about their specific needs, preferences, and technological skills to prepare adequate accommodations tailored for each participant[1,2,3,4]<br>- (Before scheduled PD session) Emailed and texted reminders multiple times before scheduled session of both the session time/date and required study materials[2] |
| **Throughout the PD session** | - An experienced TBI rehabilitation professional leading the session focusing on providing communication and cognitive supports, and other members offering (primarily) technological and input support[1,2,3,4]<br>- Provided sufficient time for tasks and increased frequency and length of breaks[2]<br>- Provided clear and repeated instructions[2]<br>- Provided multiple modalities for information (e.g., via images, text, verbal explanations)[1,4]<br>- Provided documentation of participants' responses in real time (e.g., digital sticky notes)[1,2]<br>- Provided example of a response to study prompts when necessary[2]<br>- Provided step-by-step tutorial on using Zoom features (e.g., screenshare) when necessary[3]<br>- Provided remote technical support via Zoom remote control when necessary[3] |
| **Introduction and Exploration** | - Rapport building: detailed introduction of the research team and sharing TBI story[1,2]<br>- Used digital sticky notes to keep track of and organize participant's thoughts[1,2] |
| **Ideation** | - Prepared brainstorming activity of a familiar event to simulate rapid ideation (e.g., What are the different uses of a toothbrush?)[2]<br>- Took turns generating ideas or shared random images to spark thinking when necessary[2]<br>- Provided multiple ways to input responses for ideation (e.g., writing on paper, verbal, online board)[4]<br>- Used digital sticky notes to keep track of and organize participant's ideas[1,2] |
| **Sketching** | - Showed examples of design ideas or took turns discussing solutions to sketch when necessary[2]<br>- Provided multiple modalities for sketching (e.g., drawing on paper, verbal, online board)[4] |

*3.3.1 Introduction*

Before each participatory design session, one participant and two researchers joined the Zoom meeting room. One of the researchers with more than ten years of experience as a speech-language pathologist working with individuals with TBI was the facilitator for the session. Another researcher was primarily responsible for taking notes (documentation and digitization of participants' responses in real time) and providing technology and input support for participants when necessary. Once participants entered the meeting room, the facilitator explained the goal and procedures of the study. The facilitator described the participant as "expert" in designing the social media sites for people with TBI and the researchers as "students" in this session. After obtaining consent for recording the Zoom meeting, the facilitator informed participants that they could take a break or stop the session at any time. Then, we introduced ourselves to them in detail, in particular, noting the facilitator had experience working with the TBI community for many years and another researcher would be responsible for providing various kinds of support as needed. Through this we sought to ensure that participants felt comfortable sharing their TBI experiences and engaging in PD activities. We then asked participants to share freely about their own TBI story (e.g., when, and how their injury happened, post-injury challenges, life changes post-TBI) as much as they wanted to disclose, so that researchers could start establishing rapport with them and gain a better understanding of their Facebook experiences in relation to their TBI.

*3.3.2 Exploration Stage*

In the exploration stage, the participant engaged in an in-depth interview regarding their Facebook use and specific challenges they encountered on Facebook. Additionally, we asked if and how their Facebook use had changed after their injury. Participants then engaged in a think-aloud activity, sharing their Facebook screen via Zoom screen share while navigating the site and describing their thoughts in real time. While participants freely shared their experiences and challenges in using social media with the facilitator, another researcher digitally captured and grouped the main points they addressed in real time as sticky notes (**Figure 1**). After the think-aloud activity was complete, researchers shared the sticky notes with participants to confirm that they accurately reflected participants' comments.

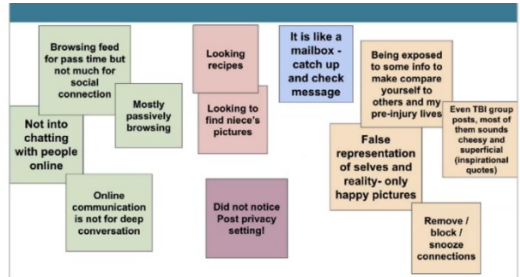

Figure 1: Digital Sticky Notes from the Exploration Stage (from P6's PD session)

*3.3.3 Ideation Stage*

From the sticky notes from the Exploration stage, the facilitator guided participants to reflect and choose the most critical challenges they wanted to address in this PD session. In the first two sessions, participants were initially asked to generate "*How might we*" questions, a tool that is widely used to define design goals for ideation [51,79]. The first two participants stated that those questions were too abstract and difficult to address. Thus, we changed the goal identification activity to a "fill-in-the-blanks" format (i.e., "I want Facebook to (be) ______ so that I can ________"), which was well-received by the remainder of the participants. Once participants created one or two design goals based on their challenges, the facilitator introduced general brainstorming rules such as 'defer judgment' and 'go for quantity.' For each design goal, the researcher asked participants to generate 4-5 ideas. For those who were unfamiliar with brainstorming activities, we provided timed pre-brainstorming activities (e.g., 'imagine alternative uses of a toothbrush'). Participants could also choose from various methods of response, such as handwriting, typing, or just verbalizing ideas with the notetaker taking notes. After brainstorming, the note-taker digitally captured and shared participants' ideas with them so that they could explain their ideas one by one.

*3.3.4 Sketching Stage*

After participants generated ideas, the facilitator asked participants to choose a couple of their favorite ideas and further develop and sketch them. We explained that sketches could be anything from conceptual drawings to a textual description of the system's key features and provided examples to make the sketching activity less burdensome. Participants could complete their drawings with paper and pen or using an online drawing board (Jamboard [117]) or have the researcher draw it out for them with the participant's verbal guidance. The note-taker digitally captured each completed sketch, and the facilitator prompted participants to elaborate on their ideas.

*3.3.5 Wrap-up*

As a final step, the researcher asked participants to evaluate their design solutions and provide feedback for the design session. Participants then filled out a post-session questionnaire with demographic information, TBI-related information and symptoms, social media use, and feedback on the participatory design workshop. Participants were compensated with $50 gift cards.

## 3.4 Data Analysis

All PD sessions were screen-recorded using Zoom and recordings were transcribed by graduate student research assistants. All participants' brainstormed ideas and sketches were digitally captured during the session and integrated into the transcription of each participant. Based on the structure of our PD procedure, transcripts were divided into two parts: 1) challenges faced by social media users with TBI (mainly from the Introduction and Exploration stage), and 2) design ideas for users with TBI (primarily from Ideation and Sketch stages). Using the thematic analysis method [82], three researchers read each transcript multiple times and generated initial codes individually, focusing on how each solution idea related to the different types of TBI-related challenges and barriers to Facebook use. Next, we iteratively created common themes by identifying frequently occurring codes across researchers, which allowed us to develop higher-level themes. Higher-level themes were then divided into the four groups that we describe in the next section.

## 4 FINDINGS

All participants completed all the procedures. While some required multiple breaks due to fatigue, our PD process was well received by our participants overall. Participants said the process was accommodating and encouraging: "*I felt heard and supported and encouraged to participate*" (P9). Others also said they felt empowered and proud of helping envision technology to improve social media experiences for other individuals with TBI and themselves: "*I want to say our reality is still valid in this world even though it's not necessarily perfect. [...] I'm fascinated by the fact that we can still contribute to society*" (P10). In the following sections, we first report participants' comments on their overall Facebook experiences and current usage patterns, as a context for the main results. We then present four sets of challenges that individuals with TBI reported in the PD sessions, followed by solution ideas proposed by participants.

### 4.1 Returning to Facebook after Injuries

Nine of the 10 participants had a Facebook account prior to their injury. Most of those with preinjury accounts said they significantly reduced their Facebook use in their early days after injury. In particular, participants said that their sensory (e.g., light sensitivity) and cognitive symptoms made it difficult for them to be on digital screens for an extended period of time. P10 shared, "*Due to physical pain, cognitive, emotional difficulties, I limited my use of social media.*" They mentioned therapy and rehabilitation training they received at the time of diagnosis helped them to regain basic sensory and cognitive functions (e.g., vision, reading, speaking) to some extent. However, as presented in Table 1, participants reported sustained sensory, cognitive, and psychological symptoms related to their TBI, but most mentioned that they could not afford extra therapy: "*I can't even get targeted therapy, because all the therapy out there is so expensive*." (P2).

Two participants stated that they had deactivated or deleted their Facebook accounts for a while after their injuries. P8 explained: "*I got rid of [my account] because it was a time suck, right?*" P1 said he deactivated his Facebook because he had felt humiliated due to his life changes brought about by TBI and wanted to withdraw from any kinds of social interactions: "*I deactivated it and kept it that way for a long time. I didn't want to have any contact with people. I wanted to just hide everything about TBI. I was probably embarrassed.*"

At the time of the study, all participants reported that they had returned to Facebook and logged on at least a few times a week. A few of them even reported browsing it several times a day: "*I'm checking it every hour; I'd say at least every 2 hours*" (P5). In the think-aloud session, all participants appeared familiar with Facebook and used it competently. As a whole, participants generally agreed that social media had been a valuable resource for maintaining social connections and searching for information, which was why they still maintained Facebook accounts in spite of challenges described below. For example, some recognized social media's unique benefits for helping them cope with memory impairments, by allowing them to retrieve information about others (e.g., birthdays) and also themselves. P9 stated, "*Most of the time, I'm using it almost like a reference point.*" Several participants mentioned using Facebook to search information about TBI. P3, who had not been very active on social media before injury, said that she gradually came to accept that social media could be helpful for accessing TBI-related resources: "*[Facebook] didn't even occur to me as a resource [before injury]. I've explored all kinds of stuff about TBI via Facebook. I think a lot of people with TBIs end up doing a lot of research via Facebook.*" Some participants also appreciated that they could connect with other TBI survivors in support groups on Facebook: "*I'm in a number of groups, which has been very helpful. It's nice to have a support network with people who understand what you're going through.*" (P2).

Most participants reported experiencing changes in their Facebook usage following TBI. Many participants mentioned that they reduced the amount of time spent on social media after injury mostly due to sustained sensory and cognitive symptoms. For example, P9 mentioned "*Definitely my Facebook usage has changed [since injury]. I had a really hard time looking at screens--because of the light. I mean even just a few minutes, it costs a lot of discomfort.*" On the other hand, a few reported increased uses of social media after injury as a result of boredom caused by reduced social participation: "*I find I am on it more especially when recovering from a TBI because I get bored as I am not doing anything.*" (P2).

Besides the changes in the amount of time they spent on Facebook, they also reported changes in other areas. Several participants said they were attempting to limit Facebook connections to those who were closest to them after injury: "*[When creating a new Facebook account after deleting their pre-TBI account], we decided to have a really private social media presence*" (P8). Also, some reported preferring maintaining limited social presence on Facebook by using it more passively than actively (e.g., consuming rather than posting). P5 shared, "*I don't have any intentional posting on my wall. I don't think I'll ever do that.*" In addition, a number of participants reported limiting their Facebook uses only for a specific purpose (e.g., marketplace, business networking), as P4 stated: "*I do only short things on Facebook. I have tried my best not to be on it anymore. I used Facebook for a specific reason.*" As such, while all participants reported frequently being on Facebook and acknowledged its value, they experienced some level of changes and limitations in their social media use after injuries.

### 4.2 Interface, Feature, and Information Overload

Participants reported lingering physical and sensory symptoms associated with TBI including vision problems (e.g., sensitivity to light, blurred vision), headache, and intense fatigue, which made using digital screens difficult for them. P9 stated, "*I had a lot of headaches and light sensitivity. I had a really hard time focusing, mostly ongoing,*" indicating that symptoms continued years after the injury. Participants also reported symptoms related to cognitive impairment, such as poor short-term memory, lack of attention, and difficulty sequencing and understanding information.

TBI-related symptoms negatively impacted participants' Facebook use. Over half of the participants reported feeling "overloaded" when using Facebook. Some participants complained about "interface overload," saying the current Facebook interface was too confusing and complicated. For example, P8 expressed his frustration with the Facebook interface: "*It's pretty busy, right? It would be really tempting to get super distracted.*" P9 shared her struggles with digesting the interface elements and switching and maintaining her focus on each part of the interface: "*My biggest barriers is to reintegrate all the things on the [screen]. I don't know how much I was trying to intently focus on something small. I don't know who I'm looking at when I'm on it. And I don't even remember why I'm trying to use it.*" There was a greater overload in navigating interface structures on Facebook when participants returned to Facebook post-TBI. P1, for example, shared that when he reactivated Facebook after a while post-injury: "*I couldn't find [messenger]. I just wanted to send a quick chat to someone. But I didn't want it up after because I could never find out how to do it*".

Participants also mentioned "feature overload", reporting difficulty navigating different features on Facebook. P10 explained that being on Facebook required her to do "*multitasking*", as she was constantly distracted by different features on Facebook: "*I could do one thing, but something comes up that I need to do later. I couldn't do multitasking.*" P9 also expressed frustration with relating how different parts of the features are linked together: "*It feels somewhat disconnected. You have to mentally follow a lot of steps to get to understand why you are looking at the thing you're looking at.*"

In particular, several participants reported feeling extremely uncomfortable and confused whenever Facebook updated its interface and features. After TBI, they said they had tried hard to rebuild a new "mental model" and routine of how to access and use Facebook's features, repeatedly trying each feature to re-learn how it worked. P4 said, "*By tapping on my own profile picture [icon], I can find [the search box] at the top left, and that's where I can find out where that is*", showing how they recognized the location of certain features and accessed them. However, when those interfaces and features updated, as was often the case, participants reported feeling helpless as they felt forced to adapt: *"There's often annoying things. Like when they make changes, you just need to learn to live with them, right?"* (P3). P2 expressed frustration with the unexpected feature changes and lack of response from Facebook: "*When I report the issue, I get no response. This makes me feel angry, frustrated, unvalued, not heard, downright mad. I feel like being left in the dark.*"

Some participants reported experiencing "information overload" due to the amount of information they were exposed to through Facebook. P1 pointed out that, "*it might be difficult to figure out stuff intuitively especially for people with brain injuries. The easiest way to put it is just 'information overload.' Like for people with a TBI, there's just too much information*." Several participants also reported experiencing extra burdens when dealing with notifications, due to their tendency to miss them as well as to become distracted by them. Therefore, P3 said she deliberately tried to limit her notifications because she *"wanted to reduce [cognitive] intake.*"

Overall, there was a general feeling of overload when it came to the interface, features, and amount of information on Facebook, making it difficult for participants to search for information or locate certain features they wanted. Many individuals with TBI reported having difficulty with tracking, switching, and maintaining attention, and often felt lost: *"I have had lots of difficulties with tracking, like 'where am I?' and understanding and processing. So, I had to go back and read again and forget it. I assume other TBI survivors also [experience similar things]"* (P10). Some responded very emotionally when they faced these challenges, as reflected in P1's account: "*What definitely bothers people with TBIs is that there's so much stuff and hard to find what you want. [For me] it's hard to manage anger when we can't find something.*"

*4.2.1 Solutions*

During the ideation and sketch stages, five participants chose to develop their ideas for addressing challenges with the interface and features, and information overload. For instance, a design goal set by P1 was "*I want to minimize Facebook so that I can use it more confidently and efficiently.*" P9's goal was "*I want Facebook to be easier so that I can understand what is happening.*" Participants expressed a strong desire to make Facebook simpler, more customizable, and more intuitive so they could easily locate features and information they were seeking.

Some participants, particularly those who found Facebook's current interface confusing and overwhelming, proposed an interactive guide as a way to help them identify what to look at and where to focus their attention. P4 shared his idea of an interactive guide that could detect users' point of gaze and help them switch focus to the feature they want to access: *"It would be nice if there were something that would interact with me and say 'hey, check this out right off to the side here', and you will get something."* Similarly, P1 envisioned an avatar-like interactive guide: "*[The avatar can say] start here, then go here. It would like a guide to make it clearer to indicate 'things you're looking at' and 'things you should look at.'*"

Aside from wanting to better navigate the current Facebook interface as it is, participants also wanted an interface that could be customized to make it simpler and more personalized. P4 suggested customization options for resizing and rearranging different sections of the interface: "*I want some customizable options, where I can dictate where all my stuff is. I would like to have my own changes as in World of Warcraft (WoW) where I can direct and drag the chat log and put it anywhere, any size I want.*" Similarly, P1 suggested a customizable interface that only included features and menus he frequently used. In his sketch (**Figure 2**), he restructured the Facebook interface to contain only the big buttons to features he frequently used (e.g., chat, news, search bar). Additionally, P10 proposed an option to customize the color, contrast, size, and spacing of fonts, because many TBI survivors, including P10, had vision limitations and trouble processing texts on social media: "*I want Facebook to allow me to choose the right font, color, and size to communicate my message more effectively.*"

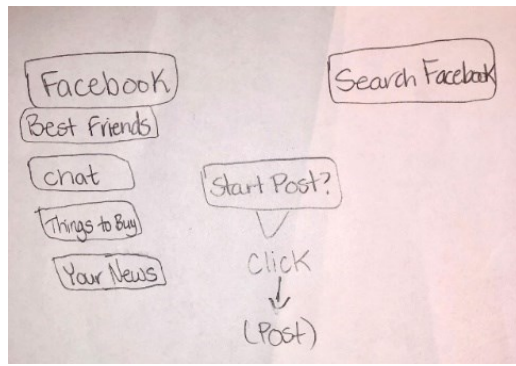

Figure 2: Customizable interface that contains only frequently used features and menus (P1)

Some requested information prioritization mechanisms for their newsfeed, so that they could allocate their limited cognitive resource to more relevant and important information to deal with information overload. For example, P3 emphasized that as many TBI survivors "*got limited [cognitive] bandwidth*" so Facebook needed to "*figure out to prioritize [information]*". Similarly, P4, who wanted to see posts related to his interests (e.g., paintball), suggested a "*ranking system*" that displayed posts with high relevance at the top. Also, P1 suggested a separate feed for "*best friends*" that would show only the posts from his close friends.

### 4.3 Facebook as a source of negativity and social comparisons

Participants reported a range of emotional symptoms after TBI, including difficulty controlling their emotions. For example, P3 described her emotions as "*all over the place.*" Some participants experienced amplified emotions, as they unpredictably became irritable or anxious: "*I'm extremely irritable. Like I'll cry or yell for absolutely no reason*" (P2).

Emotional lability was manifested in the way participants used Facebook, particularly when browsing the newsfeed, where diverse topics and opinions of posts were displayed. Participants stated that being exposed to unanticipated and unwanted types of posts often triggered an emotional reaction. P8 even referred to the overall Facebook experience as a "*triggering activity.*" Such triggering events usually referred to participants' negative reactions to being exposed to certain types of posts. For instance, P3 expressed discomfort after viewing a horror story post in one of her groups, saying: "*It's just too much for people with brain injury.*" Several participants also mentioned that posts about political, proactive, or polarizing topics often triggered negative feelings. Thus, they wished to avoid being exposed to such posts as much as possible, as reflected in P8's account: "*Anything political, racially provocative. I don't go to Facebook to read that stuff. [...] Facebook is not where I go for that.*"

A few participants reported being offended and annoyed by algorithmic recommendations for sponsored posts and friends. P8 expressed frustration when he encountered a sponsored ad while browsing his feed: *"I don't know why it's there. I want to mute crap like this. It's offensive to me and makes me so uncomfortable."* Additionally, he added his thoughts on friends' suggestions: "*Even though our focus is on really close friends and family, we still get these suggestions for just, frankly, random people. I'll tell [Facebook] when I want to add somebody. [Facebook] is not smarter than me.*" Interestingly, P3 expressed ambivalence about algorithmic suggestions. When she repeatedly received suggestions for TBI groups, she felt it was too "*freaky*" that Facebook algorithm knew she had a TBI. At the same time, she saw value in the Facebook algorithm knowing about her injury, because it could provide useful resources: "*When I see [TBI group recommendation] appear, I felt that it was just too freaky. But I don't know, maybe there needs to be some way that that information can reach survivors more.*"

Many participants also mentioned that negative feelings were triggered by posts from their Facebook connections. Some participants believed that people typically post only sugary representations of themselves and their lives on social media and held very negative opinions about that. For example, P5 shared: "*I just think of Facebook as being pretty superficial and fake, or [presenting] inaccurate depictions of people's life.*" They pointed out that such embellished descriptions left many people feeling inadequate and ashamed of their own lives, as in P1's account: "*Everyone puts their best stuff to make themselves look good, which kind of makes everyone simultaneously think they're worse than other people, too.*"

As reflected in P1's account above, many participants stated that they often compared themselves with non-injured and pre-injury peers through posts about their life achievements, such as marriage, children, and career success, which made them feel even more inadequate, because their own lives had been ravaged by the injury. For example, P6 stated that looking at the posts from her pre-injury friends caused her to compare herself to them and ultimately made her see herself as 'not normal': "*A lot of times, there's just so much going on in other people's lives. I think to myself: 'if I didn't have my brain injury, I could be experiencing all these things too.' I don't really think Facebook is healthy. I mean that's gonna overwhelm you with sadness. I hope I can see more accurate representations of life so I can feel normal.*"

Many participants tried to control what they would be exposed to and notified about on Facebook, to reduce their chances of seeing "trigger" posts. As an example, P8 said that he actively hid posts that were "*triggering*" and unfollowed pages and people who frequently posted unpleasant content: "*I hide all of their posts. At least 700 are hidden.*" However, exercising such control was a challenge for many participants, because it was not intuitive for them to figure out how to do it. For example, P3 described the challenges she faced when configuring her notification settings to only receive notifications she wanted: "*It's not necessarily easy to remove some and only get [certain] notifications.*" Thus, instead of systematically curating Facebook settings, some participants tended to skim their feeds: "*A lot of people are really bothered by the drama and the negativity, but I just don't let that [happen]. If I don't like it, I just keep scrolling*" (P7).

*4.3.1 Solutions*

In the ideation and sketch stages, four participants chose to develop ideas to reduce emotionally triggering experiences while on Facebook. Overall, these participants envisioned the interface giving them more control over the content to which they would be exposed. For example, P9 set her design goal as, "*I want Facebook [to] allow me to determine what you are presented with*." As a result, P9 desired a completely configurable and self-directed feed system instead of algorithmically curated feeds. According to his design (**Figure 3**), the user can control the type of source and posts they would be shown in their newsfeed: "*I want something that feels really centralized around [my choice]. Like when I am clicking, this is what I am choosing and seeing*."

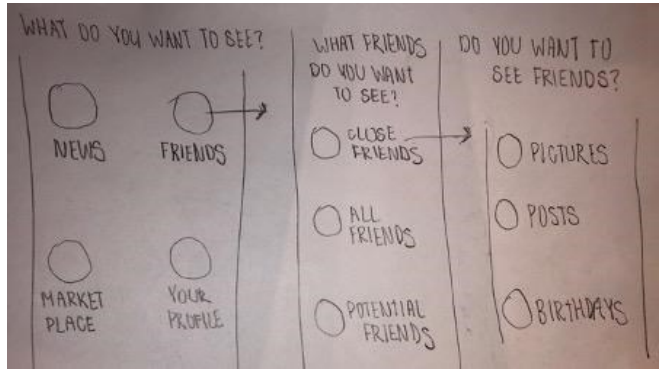

Figure 3: Configurable feed system that users narrow down to the kinds of information they want to see in their feed (P9)

Other participants suggested feed filters, such as systems that allow them to specify and eliminate emotionally triggering posts. P5 envisioned a system that would let him skip posts related to potentially triggering topics: "*Let's skip pandemic-related posts and election-related posts. So, it's kinda dependent on the current time we're in. I'd probably be annoyed right now about posts related to inflation*." Similarly, P8's suggestion was to add a "*safe filter*" (**Figure 4**) to Facebook feeds so users would be able to filter out emotionally triggering and negative posts: "*[In the] welcome screen, the first one would set a filter [for] offensive, garbage posts. [Initially on the feed,] there are skulls and crossbones and some good stuff and maybe some sad faces there. All of that goes in [the filter] and what comes out is something good.*"

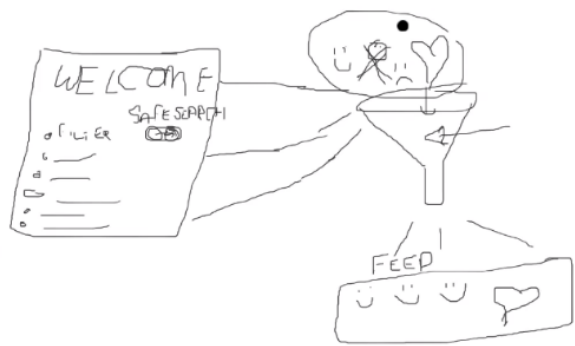

Figure 4: Safe filter to remove unwanted posts on user's feed on Facebook (P8)

### 4.4 Self-presentation Concerns Related to TBI

As a result of their brain injury, participants reported feeling forced to make some undesirable life decisions such as quitting their jobs or changing careers. For example, P8 said he had been laid off for two months after his TBI, and P2 and P3 said that they had to quit their teaching and editorial jobs, respectively. P3 recalled how disappointed she was when she was told that she could not return to work: "*My number one goal was to get back to work. So, I started working with [an occupational therapist], and she said, 'forget about the work conditioning right now.' And then, it was clear I'm not going back to work yet*." In addition, some participants felt that their personalities had changed as a result of their TBI. P2 expressed her feelings of helplessness: "*I want to know what part of my brain is being impacted. But I'm not going to know, and it just makes it so frustrating, I'm not the same person I used to be. I noticed a huge personality change.*"

As a result of changes in jobs, capacities, and personalities following TBI, participants' social identities had changed dramatically. Many participants speculated that accepting "who I am" compared to "who I used to be" was extremely painful and challenging, and an ongoing process. P10's story shows the difficulty in reconciling her past and present identities: "*I imagined the mirror. The person that I used to be was not there anymore. I was someone new, but I didn't like this new me, because I wanted to go back to be the person that I was.*"

By the same token, participants who had used Facebook before their injury had difficulty reconciling their pre- and post-TBI Facebook identities. Seeing their pre-TBI Facebook identities often led participants to feel ashamed and compare their present selves to their past selves. This comparison was illustrated by P6: "*You need to have good self-control, so you don't become obsessed with comparing yourself to what you used to be, because that's what you would do on Facebook.*" P1 shared his reflection while showing us his old Facebook posts before his injury: "*To go back on my Facebook, I was a very successful person. But then I had a brain injury. Well, people are probably [thinking] 'damn, he is doing pretty well for himself'. And then probably it'd be a little embarrassing [to let them know] that I was the kid who also had to hit rock bottom.*" Then, he added "*If I put my new identity, which is actually like a dramatic, much different identity than before. There could be a time when I'd want people to know about my story, I suppose. I'm just not there yet.*"

As reflected in P1's account, many participants believed that their identity reconstruction process was still ongoing, so they were not ready to disclose their TBI and new identity on social media. Therefore, these participants were still reluctant to publicly share their TBI or new identity on Facebook. Even though many participants proudly shared their TBI story and new achievements post-TBI with the research team (e.g., obtaining a master's degree, finding a new job path), most expressed reluctance to do so on Facebook. This reluctance was illustrated by P5, who said: *"I'd be more hesitant to post about my brain injury just because I'm self-conscious and a little more insecure. Like, if I post, what will people think?"* Overall, self-acceptance was seen as more of a personal journey than a public one, so it was something participants felt they had to do on their own, not on social media: "*Everybody's journey is in getting to self-acceptance. You're going through your TBI journey trying to understand who you are because your inner and outer self aren't really in sync. [...] It's a personal issue that you have to deal with. I mean if you don't share any of it [on social media], it is not even really there, right?*" (P3).

Instead of updating their profiles, most participants chose to maintain their Facebook profile information as it was before their injuries, or to reveal as little information as possible. P3 explained why she had been reluctant to change her work information on Facebook: "*I still haven't changed my profile just because I feel like as soon as I do say I'm not working where I was, I'm gonna have to answer questions.*" Similarly, P1, who put almost no personal information on his Facebook profile, said: "*I don't want my information there on Facebook. I don't want that many people to know at all. I* *don't want that much attention.*" As such, they felt it would be burdensome to publicly share their life changes on social media, as they thought they also had to share their TBI with others to explain their reasons for those changes.

Participants' reluctance to disclose their TBI and new identity was mainly due to a fear that others on Facebook would not fully understand TBI and their experiences. As reflected in P10's account, some did not expect that they could receive much support or understanding from uninjured peers, as they no longer had interests in common: "*There was nothing in common [with others on Facebook] anymore, because I could not do social activities while I was mostly living at home and [experiencing] physical and emotional pain.*"

In contrast, participants expressed a strong desire to closely connect with other individuals with TBI who might be able to understand each other. Therefore, many mentioned that they had tried searching for and joining TBI support groups on Facebook when they came back to Facebook after TBI. Although they found TBI support groups somewhat helpful and useful, some participants pointed out that those groups did not facilitate meaningful relationships or authentic communication among individuals with TBI. For example, P6 stated that many of the Facebook TBI groups were "*too cheesy*" and "*superficial*" because most of the shared content was inspirational quotes, which she did not find were actually helpful or relatable. Further, P9 shared that attending TBI support groups was "*contributing to the feeling of isolation because I can't find anyone that has a similar experience as me*", mainly because the groups were too large.

Interestingly, P3 expressed concern that networking with other individuals with TBI and joining TBI groups on Facebook could put their privacy and identities at risk, given that she had not publicly disclosed her TBI. She became aware of this risk when she discovered that an acquaintance was also a member of the same Facebook TBI support group to which she belonged: "*I mean that's privacy issues, [because of] overlaps of people in different groups. You also don't want people to necessarily know that you're in a group. So, then you might want to actually run back to the safety of your own feed, and screw that. And sometimes it feels like you're invading someone else's privacy by seeing where they are.*" As she did not disclose her TBI on Facebook and she was also not sure whether her acquaintance wanted her to know, she questioned whether connecting with other individuals with TBI and joining TBI support groups on Facebook was a safe option.

Some participants explored social media outlets other than Facebook, such as Instagram. In setting up their new Instagram accounts after TBI, these participants said that they felt more in control of who they could talk to and what they could say about their TBI. Because they set it up after having TBI, they could choose to be friends only with people who understood them or others with TBI who were already on Instagram. Therefore, they were more comfortable sharing their TBI stories on Instagram rather than on Facebook, where so many of their pre-injury connections were included. P10 described her Instagram experience as, "*It opens different doors. I find that through Instagram, I make better connections. [...] I post about brain injuries, so now I connect with people who live in England and Australia [who also had TBI].*"

*4.4.1 Solutions*

Overall, three participants suggested solutions related to the challenges in accepting and rebuilding their new social identity. One idea from P3 focused on creating a private and secure place where she could explore TBI-related information and connect with others who had TBI without risking their privacy or boundaries. P3 created her design goal as "*I want Facebook to have a secure and private place where I can explore TBI stories and connect with other TBI survivors.*" She envisioned a subset of Facebook that she called "*TBI world*," where she could communicate with other individuals with TBI more comfortably and explore and organize TBI-related stuff separately from her other Facebook presence. She acknowledged that social media would be a powerful tool for connecting with other people who had TBI and supporting one another. However, she was reluctant to seek help from Facebook because it would eventually lead to revealing her TBI to a broader audience: "*What their biggest issues are is finding help and people who know what we're going through. I would like to share what I want with my feed, but if I do that, I'm gonna step outside of [my boundaries]*." Thus, she imagined a private and secure Facebook environment where she could stay private: "*It would be like a little separate universe that I knew I could go into where the TBI stuff is. That's like a place where you're still on Facebook, but it's like a smaller Facebook.*" Her sketch (**Figure 5**) included a separate space where she could easily talk to other individuals with TBI and access and share TBI resources with them: "*It's like people sharing their stories and the other resources such as webinars and local place(s). It's almost like a portal to your world*."

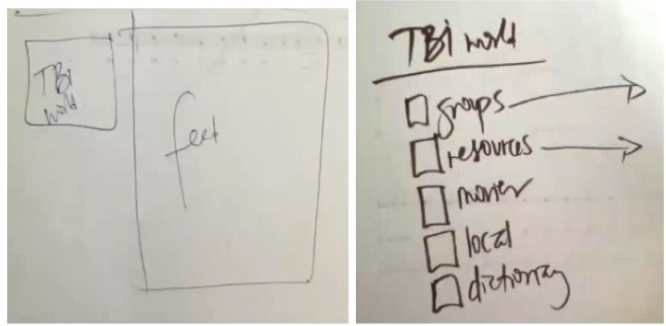

Figure 5. TBI World as a private and secure place to explore TBI-related information and connect with other individuals with TBI (P3)

P6 wished for a dedicated group for people with TBI where she could feel safe and empowered by making meaningful relationships with other individuals with TBI. Her design goal was *"I want a new version of Facebook where I can feel normal while connecting with others who also have TBI.*" She came up with the new concept of a TBI support group on Facebook (**Figure 6**): *"It would be neat if there were one dedicated group for people that had brain injuries where we could just get on and share the things that have happened and [ensure] we are still doing well in our life. [In that group] we will be with one another and can support each other in that sense, as ways of coping with life."* She emphasized the importance of interpersonal relationships among individuals with TBI, which she could not easily achieve through the Facebook support groups she had joined: *"It needs to be more personal where you actually got to know one another."* For her, most TBI groups on Facebook felt like a "*big town*" while she preferred one that felt "*more like a small town*": "*This is not a good comparison, but I think it should be more like a small-town vibe instead of a big city vibe.*"

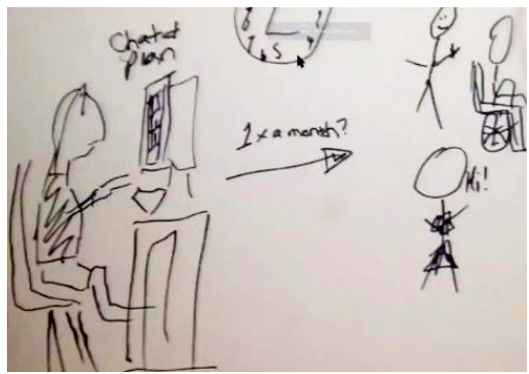

Figure 6: Smaller support groups on Facebook to connect with others with TBI more closely (P6)

To develop close interpersonal relationships, she thought the group size needed to be "*manageable.*" In that way, she believed that members could share their little daily moments to develop a lasting relationship and support each other. She also indicated that some level of group moderation would be necessary, such as scheduling regular meetups, encouraging discussion, and ensuring a safe and positive environment.

**4.5 Reduced Perceived Confidence in Communication**

TBI results in cognitive impairments, such as difficulty concentrating and organizing thoughts, all of which can affect language comprehension and production capabilities. Several participants reported difficulties understanding meanings in sentences while listening to others and reading texts, while some noted gradual improvements over time. P9 expressed her frustration when she realized she could no longer read as effectively as before: "*I almost got lost in focusing and figuring out how this [sentence] ties back to others. I felt like I couldn't read. It's like living in a cloud.*" Also, when participants communicated verbally or wrote text, some still found it hard to come up with and connect words and formulate coherent sentences: "*I felt these words aren't mine anymore. I do find myself kinda being at a loss for not knowing what and how to say.*" (P4).

Because of their communication challenges, some participants stated that they became very self-conscious when engaging in social situations, as indicated in P5's account: "*Being self-conscious about making mistakes and caring about what other people think about me can result in keeping you from doing what I'm looking to do.*" In particular, several participants reported feeling overly conscious about behaving inappropriately in social settings. P9 shared that *"Right after the accident, I remember feeling ashamed because I didn't think what I said was inappropriate. To me, I was telling the truth, or I was just saying something honestly, but it hurt people's feelings."*

Participants' communication challenges could lead to reduced confidence in communicating with others on social media. For example, P5, who decided to use Facebook for career networking, had met many new people on that platform, and sometimes he had to ask them something through messaging. He noted that he had become overly self-conscious whenever he tried to message someone: "*There's a feeling of anxiety of wondering what that person thinks of me such as 'am I asking too much?' or 'is this something that most people would do?'*" Especially on Facebook, which is still heavily text-based, some participants found it difficult to convey their intentions accurately via text messages. P5 explained: "*I mean it's challenging over text. I'm concerned about how the messages are being perceived by the receiver.*"

While all participants were technically capable of reading and viewing text, some noted the challenge of understanding others' intentions in text messages: "*For any sort of text communication, you don't have the advantage of seeing somebody's face and hearing the inflection in their voice. So, things are easily misinterpreted. So, I have to remind myself of that constantly*" (P7). Further, P9 mentioned that it was especially challenging for her to understand the "*nuances*" of others' messages: "*I have a hard time understanding. Like, I might miss the joke and I have a hard time understanding sarcasm.*"

*4.5.1 Solutions*

Five participants chose to create ideas for regaining their confidence in communication. One line of ideas focused on supports for language comprehension and production. They wished for ways to identify the meanings of certain words or some portions of sentences that did not come to them clearly. P9, for instance, hoped for a built-in feature that would read out parts of texts that were unclear to her. She thought the option of hearing unclear words would help her "*reaffirm*" what she saw. P4 suggested that displaying word meanings visually would be helpful for when he occasionally ran into words whose meanings he could not recall: *"Without a doubt, [I hope] something that you just click it and then visually tell you what it is."*

Some participants wished for more direct help composing messages, particularly tools that could help them articulate their thoughts and then find and connect words to make their messages more organized. For example, P10 envisioned a writing aid that could automatically compose messages using a few keywords: "*So you have a vague idea, but not sure how to transpose that into sentences that communicate properly or efficiently. So, you toss out some topics that you want to address, just word levels. And the system compresses and connects, and then it makes a coherent sentence. And then you can just accept or dismiss or redo it.*" P4's idea was a text-prediction-like system that could help find the words to fill in the gaps between words that he had already typed: "*In case you have a word that you're struggling with, you can put a little Asterix mark [as a placeholder] then something tells me these are all the words [you can include].*"

P9 proposed a "*pre-loaded comments*" feature designed to respond to others' posts in a relevant and appropriate way using auto-generated responses. In her design sketch (**Figure 7**), the system first describes what the post is about and shows what the typical response types would be (*"respond", "question", and "compliment"*). Then, users can choose between pre-loaded comments categorized by different purposes (e.g*., "Did you take this picture?" "Looks beautiful," "You are adventurous"*). She came up with this idea thinking back to a time when she had unintentionally offended a friend with her reply to the friend's post: *"I accidentally insulted a friend's baby, but I didn't know [at that time]. I was trying to say that your baby is really cute, but I think I described it in some way that was actually not that great. I said, 'it looked like a squirrel.' You're not supposed to say that."*

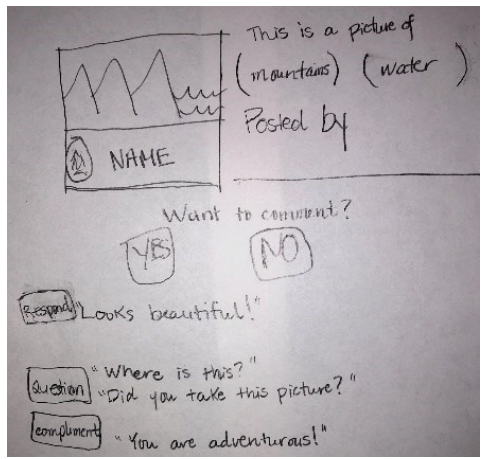

Figure 7: Pre-loaded comments to help users appropriately respond to others' posts (P9)

P5 came up with an idea to help him be more confident in communication, rather than worrying about how to compose messages and how others perceived him. His proposed solution was automatic prompts that would constantly remind him about the positive and proactive attitude he would like to have. He created lists of sentences that would cue him during social interactions, for example *"recipient does not have bad intentions.*" Also, he created checklists for "*conveying good intentions through messaging.*" Finally, he added the Bible verse "*Not letting any unwholesome talk come out of your mouth, but only what is helpful for building others up.",* to use as a reminder whenever he felt he was "*being less inhibited with my mouth*." To internalize these attitudes, he believed it would be important to repeat them, so he came up with the idea of an automated reminder system: *"If I just have something that is gonna come across my screen regardless of whether I remembered to look at it or not, that'd be helpful."*

## 5 DISCUSSION

In our findings, we first discovered that participants did recognize the benefits that social media provided, which had been examined in prior studies in terms of information access [1,18,100,114], memory aids [109], and network and relationship building [1,18,32,55,72,100]. Initially after injury, participants restricted their use of social media due to basic sensory and cognitive problems, but after a while, they resumed accessing Facebook regularly, which corroborated the previous findings [42,72]. The majority of them, however, continued to experience various TBI symptoms which had changed the way they used and perceived social media after injury. The changes they reported in their Facebook usage did not just include (primarily reduced) time spent on Facebook, but also voluntary limiting of their social connections, presence, and self-presentation activity in their Facebook use. In fact, many reported limiting their Facebook presence and social interaction and being reluctant to actively connect with a wide range of people and sharing personal updates, particularly about sustaining, and living with TBI. Previous literature has highlighted that individuals with TBI used social media as a convenient method for compensating for their social isolation and reduced social opportunities [1,32,55,100]. Our findings, however, suggest that individuals with TBI could not fully capitalize on the opportunities for connection offered by social media. This also echoes Morrow and colleagues' recent findings that people with TBI use social media as much as their uninjured peers and see its value for social participation, but they might not be able to fully take advantage of its social benefits [72].

We further characterized the challenges associated with the changes in Facebook use after injury. Four major challenges were identified from participants' lived social media experience: (1) interface, feature, and information overload, (2) triggering experience, (3) self-presentation and privacy concerns related to TBI, and (4) reduced perceived confidence in communication in a social media context. Some of these challenges (1,2) are often seen as universal drawbacks of social media use among uninjured populations. For instance, as many social media sites constantly update and add new features, the interfaces can become more complex and overwhelming, which can lead to information overload and discontinued use of social media [43,115]. Also, negative emotions triggered by being exposed to a constant stream of negative news and content and social comparison to peers have been commonly reported across many prior studies [96,99,104,105]. As such, participants' general perceptions of social media as a source of negativity and social comparison are remarkably similar to the feelings and descriptions reported by non-injured individuals in the literature [4,70]. This suggests that general drawbacks of social media can affect anyone on social media, but individuals with TBI can be particularly vulnerable since these challenges can be amplified with their TBI-related symptoms, including their impaired self-regulation and emotional regulation capacities [53,73,74].

Moreover, other kinds of challenges (3,4) have been documented in the literature on TBI. For example, avoidance of disclosing TBI to others due to fear of being judged or treated differently [87]; altered sense of identity [10,112]; and reduced communication competence [67] and confidence in social interaction [91,101] have been documented in the literature on TBI. Additionally, in studies of people with other cognitive impairments, such as dementia or stroke, similar challenges have been reported [38,40,74]. However, the focus of these studies did not directly address how these tendencies may be reflected in social media use by individuals with TBI.

As a whole, our findings suggest that even TBI survivors with relatively high technology literacy, such as our participant group, could be affected by many of the universal consequences of social media use, with challenges likely exacerbated by sustained TBI-related symptoms (e.g., sensory, cognitive, and emotional). Moreover, this study extends prior TBI literature by showing how self-presentation and self-disclosure concerns, as well as reduced confidence in communication affected their perception and use of social media.

Overall, these results point to the important design direction for future social media accessibility and assistive tools for individuals with TBI. Currently, social media accessibility tools primarily focus on sensory level support (primarily vision) [45,110]. However, we found that cognitive, emotional, communication impairments and altered sense of identity also posed great barriers, which need to be addressed to allow individuals with TBI to fully take advantage of potential benefits of social media.

To address multifaceted challenges identified from their lived social media experience, participants actively generated novel and important solutions. The types of their solutions ideas are mapped in **Figure 8** with associated TBI symptoms and challenges. Based on the ideas and sketches for addressing their own challenges, we present design implications to reduce barriers for people with TBI to capitalize the potential benefits of social media. We do not intend to represent the general views of all individuals with TBI especially as TBI symptoms experienced by individuals vary significantly based on the type, severity, and location of the injury in the brain [78]. However, we believe that participants' ideas offer a useful lens for contemplating what is currently lacking and where improvements can be made. Finally, we hope that the design implications of our findings could inform the accessibility improvements for broader populations that face similar challenge.

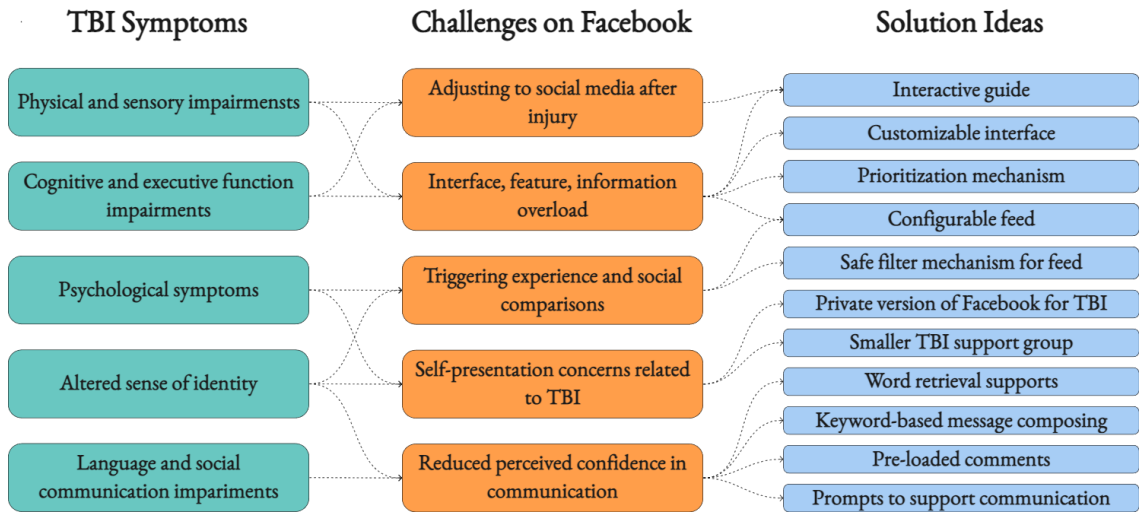

Figure 8: Thematic analysis map of findings

### 5.1 Support for Adjusting to Social Media Platforms after Brain Injuries

A common chronic symptom of TBI is limitations in sensorimotor function [60], particularly in vision (e.g., light sensitivity, blurred vision) [90]. A variety of technological aids exist to address these symptoms. Many mobile and PC operating systems and social media platforms allow users to adjust screen brightness, contrast, and blue light filter settings (e.g., [64,69]) in addition to providing screen readers [65]. However, we observed that many participants did not use such options during the think-aloud sessions, which was consistent with evidence that individuals with TBI might lack training in locating and using appropriate assistive technologies for vision and sensorimotor function [21,109]. Therefore, it will be necessary that rehabilitation professionals introduce existing tools for sensory support and guide to individuals with TBI on how to use them. In particular, many participants indicated that they could not receive additional therapy outside of the time of their diagnosis, which highlights the critical role rehabilitation professionals play in early stages of adjusting to social media by recommending appropriate tools based on each survivor's symptoms and needs.

More common than sensorimotor challenges were challenges related to cognitive impairments. Many participants reported cognitive overload when attempting to navigate the complex Facebook interface, including challenges finding and focusing on desired content, and ignoring irrelevant information and distractions. These challenges were not a function of lack of experience: most of our participants had used social media platforms for many years before their injuries, so they were already familiar with their functions. However, most TBI training programs focus on teaching basic social media features themselves [23,56].

Instead, when trying to readjust to social media sites, participants requested assistance to compensate for impairments in executive functions such as effective switching of attention, and planning and monitoring of social media activities. An idea from several participants was to create an interactive guide where users can tell the system what they are looking for and the system will show them where it is located. As this idea came from participants' challenges in directing or shifting attention to necessary interface components and navigate complex interface features, gaze-based interaction techniques (e.g., [11,37]) can be useful in designing guide and training tools for social media navigation. Further, the combination of gaze and speech interaction would be an effective approach for them because of its resemblance to face-to-face guidance sessions [39] that current social media training programs do not offer much yet.

Our results also suggest that social media sites should be cautious when updating their interfaces. Upon returning to social media, many participants reported that they tried hard to re-adjust to the interface and updates of social media platforms through trial and error. Some reported feeling undervalued, helpless and under pressure to adapt to platform changes. Therefore, social media designers will need to pay attention to the potential effects of even the smallest changes on people with impaired cognitive functioning and provide proper notice to those changes or options to skip an update.

**5.2 Gaining Control over Interface and Information on Social Media**

In addition to accepting and trying to re-adjust to existing social media sites as-is, participants imagined alternative social media platforms that were simpler, more flexible, and customizable. Increasing flexibility and supporting simplification and personalization have been core principles of the Web Content Accessibility Guidelines (WCAG) to cognitive accessibility for several years [107]. Participants' experiences and ideas for solutions, however, suggest that these principles have not been adequately addressed on Facebook, and potentially on other current social media platforms.

While participants' preferences regarding further customization varied, they generally expressed a desire to customize font styles as well as the size and location of interface components. Some participants wanted Facebook to have a more personalized look where only features they used most frequently remained. These findings suggest that customization and simplification support should be provided both at the interface and feature levels.

Further, many of the design solutions participants proposed revolved around gaining control over the amount and type of information displayed. Participants reported experiencing information overload due to the sheer amount of information to which they were exposed. Thus, some wished for a better priority mechanism for curating their feed to show posts of high relevance and interest. In addition, some participants reported being emotionally triggered and negatively influenced by posts about certain topics, as well as from certain connections. Overall, most of their ideas were focused on gaining 'control' so that they could filter out unwanted information from their feeds. One idea suggested a configurable and self-directed feed system that would allow users to narrow down or filter out posts they want to block, rather than content being dictated by algorithmically curated feeds. Other solutions suggested the ideas to allow users to specify keywords, topics, and emotions for posts to be filtered.

Facebook has already implemented some of the design ideas participants proposed. Facebook allows users to select up to 30 friends as their "Favorites" [66], so that their posts will be shown more frequently and in higher priority. In addition, users can unfollow someone or snooze their posts, which were the features that several participants reported using frequently. Interestingly, Facebook experimented with "Keyword snooze" in 2018 [76], which allowed users to temporarily hide posts based on keywords across the entire newsfeed. It appears they did not design this feature with enhanced accessibility in mind as they rather introduced it as an effective way to block spoilers from movies and TV shows. As of now (2022), it appears Facebook decided to integrate this into the "Hide" feature rather than implement it as a separate feature [76]. However, Facebook's Hide option might be somewhat confusing or ineffective for people with TBI who want more control over their feed, as this option is described as allowing users to "see fewer posts like this." Facebook might conclude that estimating the characteristics of posts based on their large dataset of user activities and word-embedding structures of keywords would be more efficient. Nevertheless, at least from some participants' perspective, having more control over characteristics of posts to filter content, as in Twitter's advanced muting options for words and hashtags [78], may enhance their social media experience.

Overall, our findings suggest that providing greater control over the configuration of social media interface and type and source of information needs to be understood as a crucial component for reducing negative feelings associated with social media browsing and ultimately reducing cognitive and psychological barriers to social media. As social media users, even those without injuries, also frequently experience cognitive fatigue [43,115] and simulations of negative emotions on social media [96,104,111], allowing users more control over interfaces and information would improve overall social media experience for broader user groups and help them to reap the benefits of social media.

**5.3   Securing Boundaries between Pre and Post TBI Identities**

One of the interesting findings was the tension between pre- and post-TBI social media identities, and how that affected participants' current social media use. Prior studies emphasized that social media can be beneficial for people with chronic health conditions in seeking information and support [18,42]. However, individuals must be able to actively discuss and seek information and support on their health conditions without jeopardizing their privacy and boundaries. Most of our participants, however, were still reluctant to disclose their TBI on social media or share their post-injury identity with their pre-injury connections.

Participants highlighted the difficulties in accepting their post-TBI selves, a finding frequently reported in the TBI literature [10,49,112]. Some of them wanted to come up with ideas around this issue, but none of them could directly address it during the ideation and sketch sessions. This indicates that the fundamental support on establishing new self-identity cannot be solely resolved by technological solutions but should be combined with counseling and community engagement in collaboration with rehabilitation professionals [23,24,48,81].

Yet, social media platforms still can offer support for individuals who are undergoing identity reconstruction after a TBI. Above all, it is important to prevent individuals with TBI from unintentionally revealing their condition on social media. Even when they do not publicly disclose their TBI on their feed or profile, there are still various ways to unintentionally reveal this by joining TBI support groups, following TBI advocacy pages, liking or commenting on TBI-related content. While such activities should be highly encouraged in the interest of gaining more information and support on TBI [18,42], they could result in a privacy and boundary breach. Therefore, as reflected in the "TBI world" idea (4.4.1), allowing to maintain a secure place separated from users' Facebook presence will be a critical element so that individuals with TBI can fully take advantage of potential benefit that social media can offer. A possible feasible solution may involve encouraging individuals with TBI to (temporarily) open and maintain a new social media account so that they can feely explore TBI information and connect with others who also have TBI. In this way, individuals with TBI may be able to access social media sites with less anxiety concerning self-disclosure and privacy for a while.

Further, as a result of group size and the nature of posts frequently shared within TBI social support groups on social media, several participants found it difficult to establish meaningful and supportive relationships. To facilitate collaborative navigation of the ongoing process to self-acceptance, online support group organizers may consider establishing subgroups within their groups and allocate subgroup facilitators, so that TBI survivors could build mutually supportive interpersonal relationships and share their day-to-day experience.

### 5.4 Providing both Language and Social Communication Help

Reduced confidence in communication abilities is one of the main reasons that many participants choose to be passive on social media after TBI. Participants with difficulty retrieving words imagined ways where the Facebook interface could read words out loud or visually illustrate them. Some suggested tools that automatically compose messages from a few keywords. One participant who was conscious of her social communication skills proposed "pre-loaded comments" that could be used to respond to others' posts. Upon these ideas, it was evident that both aids for language and social communication support would be necessary to promote active social interaction on social media. Furthermore, it may indicate that many individuals with TBI might be open to using automated communication aids.

In fact, some of the language aids envisioned by participants are readily available to some extent. For example, for word retrieval, there are numerous pop-up dictionary plug-ins that read out loud and provide word meanings and synonyms (e.g., [47]). Also, predictive text techniques (e.g., [5]) and AI-toolkits that generate sentences based on keywords (e.g., [17]) are also available. While most of them are not directly applicable to social media contexts yet, we can foresee that AI-powered reading and writing aids may be able to address immediate language challenges.

In contrast, there are fewer existing solutions to assist with deficits in social communication. In fact, social communication requires a complex set of cognitive, psychological, and social skills such as self-monitoring, emotional regulation, inference and expressions of social intentions as well as knowledge and understanding of communication contexts and cultural norms [13,33]. Therefore, it is crucial for individuals with TBI to work with rehabilitation professionals to regain various skills and knowledge associated with social communication.

Among the areas in which technological tools can contribute is monitoring pragmatic forces in messages such as sentiment [59], politeness [6,58] and formality [83] using text mining models. Additionally, crowdsourcing aids for figuring out other people's intentions and writing appropriate responses (e.g., [12]) could also be a viable option. Finally, as suggested by one participant, systems for reminding and tracking social goals would be useful along with the social communication skill development goals that individuals with TBI have at different stages of their recovery.

### 6   LIMITATIONS AND FUTURE DIRECTIONS

Our study had several limitations. First, while our sample included a range of ages (31 to 60 years old) and years post-TBI (3 to 42 years), all participants identified as white. Future studies should include individuals of diverse backgrounds, to include diverse social media users with TBI. In addition, remote participation necessitates that participants have technological infrastructure in place (e.g., computer, high-speed internet, computer knowledge). As a result, participation is limited to those with the proper infrastructure, which may amount to exclusion of other key perspectives. Future studies may consider additional strategies for recruitment of diverse perspectives (e.g., local community group).

Second, we also note the potential influence of TBI-related symptoms on active participation. For example, the "Zoom fatigue" experienced by many users of video conferencing platforms [7] appeared to exacerbate several participants' TBI-related cognitive fatigue in response to interview questions and specific PD tasks. Future studies should continue to explore accessible participatory design for individuals with TBI.

Finally, as our study focus was to generate solutions to address social media challenges, participant response bias is a potential limitation [34]. Participants may have been influenced by the interview questions and PD activities to focus more on the negative aspects of social media use. While we acknowledge this limitation, we believe that the ideas generated offer important design considerations to be explored in future studies.

## 7 CONCLUSION

Our study examined the challenges individuals with TBI face in accessing and using social media, as well as their ideas for solutions. Using remote participatory design methodology, we engaged 10 adults with TBI to design solutions based on Facebook, which is commonly used by people with TBI. Based on participants' challenges and solution ideas, we outline implications for facilitating an equal access to the benefits of social media for people with TBI. As computer-mediated communication has become the dominant method for social participation, our findings would inform the technology for supporting social participation for people with TBI, and potentially for people with other cognitive and communication disorders.